# U-net Based Direct-path Dominance Test for Robust Direction-of-arrival Estimation

*Hao Wang*[1,2], *Kai Chen*[1,2], *Jing Lu*[1,2]

[1]Key Laboratory of Modern Acoustics, Institute of Acoustics, Nanjing University, Nanjing, 210093, China
[2]NJU-Horizon Intelligent Audio Lab, Nanjing Institute of Advanced Artificial Intelligence, Nanjing 210014, China

`haowang@smail.nju.edu.cn, chenkai@nju.edu.cn, lujing@nju.edu.cn`

## Abstract

It has been noted that the identification of the time-frequency bins dominated by the contribution from the direct propagation of the target speaker can significantly improve the robustness of the direction-of-arrival estimation. However, the correct extraction of the direct-path sound is challenging especially in adverse environments. In this paper, a U-net based direct-path dominance test method is proposed. Exploiting the efficient segmentation capability of the U-net architecture, the direct-path information can be effectively retrieved from a dedicated multi-task neural network. Moreover, the training and inference of the neural network only need the input of a single microphone, circumventing the problem of array-structure dependence faced by common end-to-end deep learning based methods. Simulations demonstrate that significantly higher estimation accuracy can be achieved in high reverberant and low signal-to-noise ratio environments.

## 1. Introduction

The extraction of the direction-of-arrival (DOA) of the target speaker plays an important role in many acoustic signal processing applications, such as speech enhancement, robot audition and video conferencing. The commonly utilized algorithms, including the time difference of arrival (TDOA) [1], the steered response power (SRP) [2] and the subspace methods [3], suffer from considerable degradation of estimation accuracy in adverse environments with high reverberation and strong noise. It has been noted that the identification of the time-frequency (TF) bins dominated by the contribution from the direct propagation of the target speaker can significantly improve the robustness of the DOA estimation [4]. Several methods have been proposed to retrieve the direct-path information, including the coherence test [5], the direct-to-reverberation ratio (DRR) [6] based test, the direct-path dominated (DPD) test [4], [7] and its various variants [8], [9], among which the DPD-test-based methods are regarded as the state-of-the-art (SOTA) solution [9]. However, these methods are usually designed to alleviate the influence of reverberation and mild diffuse noise [8], and the effective retrieval of the direct-path information in more adverse environments is still a challenging task.

The data-driven deep learning technique, in the form and neural network with many layers, has achieved huge success in image related applications [10], and has also attracted interest in the field of audio processing [11]. The normal multi-layer perceptron model (MLP) [12], the convolutional neural network (CNN) [13], the residual network (ResNet) [14], and the convolutional recurrent neural network (CRNN) [15] have been utilized in DOA estimation, usually in an end-to-end form with the desired DOA acting directly as the training target. Despite its potential benefit of improved performance in adverse environments, this end-to-end implementation faces the challenge of generalization, i.e., the DOA estimate might be severely deviated in unseen noise scenarios. Moreover, the optimized network only serves a specific array with fixed microphone number and distribution, and it is difficult to adapt to different array structures.

In this paper, the strong segmentation ability of deep learning is utilized to extract the direct-path TF bins of the target speaker, based on which the DOA can be estimated using the common rule-based algorithms. Obviously, the processing of speech in TF domain is analogous to image processing. Motivated by the successful implementation of U-net in biomedical image segmentation [16], a multi-task U-net architecture is designed to estimate the ideal ratio masks (IRMs) of both the entire speech (including the reverberation) and the direct-path speech component simultaneously. The estimated IRMs are further utilized to refine the direct-path TF bins of the desired target speaker. The training and inference of the proposed architecture depend on the input signal of only one microphone, making it suitable to be implemented in any array configuration. The DOA is finally estimated by the common algorithms like the SRP-PHAT [2] on the extracted direct-path bins, which can alleviate the generalization problem faced by the end-to-end deep learning approach in untrained conditions.

## 2. Algorithm description

### 2.1. Signal model

In frequency domain, the signal captured by the microphone array can be written in vector form as

$$\mathbf{x}(t,f) = \mathbf{g}(f)s(t,f) + \mathbf{r}(t,f) + \mathbf{n}(t,f), \quad (1)$$

where $\mathbf{x}(t,f)$ is the captured signal vector at time index $t$ and frequency index $f$, $s(t,f)$ is the target speaker signal, $\mathbf{g}(f)$ is the direct-path transfer function vector between the desired speaker and the array, $\mathbf{r}(t,f)$ is the reverberant speech, and $\mathbf{n}(t,f)$ is the noise signal vector uncorrelated to the desired speech. Note that the direct-path component $\mathbf{g}(f)s(t,f)$ includes the most precise information of the target speaker DOA, and extraction of the direct-path TF bin can significantly improve the robustness of

DOA estimation [8]. Furthermore, to guarantee a reliable estimation and alleviate the influence of reverberation and noise, it is better to extract the direct-path TF bin satisfying

$$\|\mathbf{g}(f)s(t,f)\|_2^2 \gg \|\mathbf{r}(t,f)\|_2^2 + \|\mathbf{n}(t,f)\|_2^2.$$

### 2.2. The multi-task U-net

The U-net architecture [16] consists of a contracting encoder to analyze the whole image and a successive expanding decoder to produce a full-resolution segmentation, together with the skip connections between opposing convolution and deconvolution layers to combine low level detailed information and high level semantic information. It has been widely accepted as the SOTA solution on biomedical image segmentation [17]. The extraction of the direct-path desired speech in STFT domain amongst reverberation and interference can be regarded as a segmentation problem. Besides, the extraction of entire speech and direct-path speech signal utilize similar feature information from spectrogram, and can be seen as related tasks. Therefore, it is reasonable to establish a network structure based on U-net.

As noted in the field of speech enhancement [18], it is difficult to alleviate the influence of both reverberation and interference. Thus extracting the direct-path speech using U-net in a straightforward manner is not a proper choice. Actually, our numerous tests have demonstrated that the noise dominated bins are often mis-classified as the direct-path bins of speech. To make a more robust direct-path speech extraction, a multi-task network is proposed in this paper, which extends the original U-net architecture by adding another decoder, as shown in Fig. 1. The network aims at estimating the IRM of both the desired speech including reverberation ($IRM_s$) and the direct-path speech ($IRM_d$) simultaneously, and these can be utilized by a robust direct-path speech extraction scheme as described in Sec. 2.3.

Our proposed multi-task U-net is show in Fig. 1. Each blue box corresponds to a multi-channel feature map, and the number of channels is denoted on the top. The x-y-size is provided at the lower-left edge of the box. White boxes represent copied feature maps, concatenated with a nearby blue box which is the output of a de-convolution operation. The purple arrows denote a 2D convolution layer with a filter size of 3×3, each followed by batch normalization (BN) and exponential linear unit (ELU). The number of filters for convolution operation is the same as the channel number of the following blue box. The grey arrows denote copy operation. The red arrows denote 2×2 max-pooling operation with stride 2 for down-sampling. The green arrows denote the 2×2 de-convolution operation that halves the number of feature channels, followed by "ELU" nonlinearity. Finally the feature maps with the initial resolution are processed by a 1×1 convolution operation, followed by "Sigmoid" nonlinearity, represented as cyan arrows in Fig. 1.

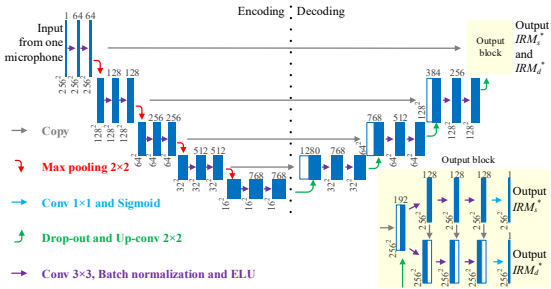

Figure 1: *Modified multi-task U-net architecture.*

Drop-out layers with rate 0.5 before each of de-convolution operation perform further implicit data augmentation and avoid overfitting. The input size is $L \times K$, with $L$ and $K$ the numbers of time frames and frequency channels, respectively. All convolutions are padded so that the shapes of the outputs, denoted as $IRM_s^*$ and $IRM_d^*$, are the same as the input.

It has been noted that the skip connections between the encoder and the decoder guarantee the propagation of the gradient flow and allow low-level information to flow directly from the high-resolution input to the high-resolution output [16]. In the output block of the multi-task U-net architecture, some additional convolution layers are used for the second task, i.e., the $IRM_d^*$ prediction, as shown in the lower right part of Fig. 1. The additional connections between the two prediction tasks transfer the feature information related to $IRM_s^*$ to the feature information related to $IRM_d^*$, aiming at improving the estimation accuracy of the latter. The input to the network is the logarithmic magnitude spectrogram of a single-channel noisy signal.

In the training stage, the desired values of $IRM_s$ and $IRM_d$, namely $IRM_s^t$ and $IRM_d^t$, are calculated as follows. The $IRM_s^t$ is defined as

$$IRM_s^t(t,f) = \frac{P_d(t,f)+P_r(t,f)}{\max\left[P_d(t,f)+P_r(t,f)+P_n(t,f),\xi_n\right]}, \quad (2)$$

in which $P_d(t,f)$, $P_r(t,f)$ and $P_n(t,f)$ denote the power of the direct-path speech, reverberant speech and noise, respectively. $\xi_n$ is a small regularization parameter for maintaining the stability of the algorithm at bins with ultra-low signal power. Similarly, the $IRM_d^t$ is defined as

$$IRM_d^t(t,f) = \frac{P_d(t,f)}{\max\left[P_d(t,f)+P_r(t,f)+P_n(t,f),\xi_n\right]}. \quad (3)$$

The cost function of the network is the summed mean squared error (MSE) loss between the estimated and desired IRM values, described as

$$L=\frac{1}{2}\mathrm{MSE}\left[IRM_s^*,IRM_s^t\right]+\frac{1}{2}\mathrm{MSE}\left[IRM_d^*,IRM_d^t\right]. \quad (4)$$

### 2.3. Robust extraction of direct-path speech

As mentioned in Sec. 2.2, it is not a proper choice to extract the direct-path speech by only using $IRM_d^*$. A small $IRM_s^*$ also indicates that the TF bin is highly likely dominated by the noise signals. Therefore, an effective refinement of the direct-path dominance test criterion is proposed by exploiting both $IRM_s^*$ and $IRM_d^*$ as the following

$$IRM_{DPD}(t,f) = \begin{cases} 0, & IRM_s^*(t,f) < IRM_0 \\ IRM_d^*(t,f), & IRM_s^*(t,f) \geq IRM_0 \end{cases}, \quad (5)$$

where $IRM_{DPD}$ denotes the refined $IRM_d^*$, and $IRM_0$ is a threshold value. It is expected that this refinement can further eliminate the influence of noise, ensuring a more robust DOA estimate. The set of TF bins passing the direct-path speech domination test is defined as

$$\Pi = \{(t,f): IRM_{DPD}(t,f) > TH\}, \quad (6)$$

where $TH$ is a threshold value.

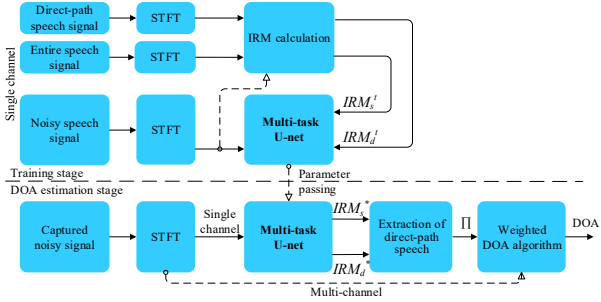

Figure 2: *Overview of the system.*

### 2.4. Overview of the whole system

The overview of the proposed DOA system is illustrated in Fig. 2. Following short-time Fourier transform (STFT), the multi-task U-net architecture is utilized to estimate both $IRM_s^*$ and $IRM_d^*$ at the same time. In the training stage, the target values including $IRM_s^t$ and $IRM_d^t$ are calculated from the single-channel signals. In the stage of DOA estimation, the estimated mask outputs are further utilized to refine the direct-path TF bins of the speech. Finally, the target speaker DOA is estimated by the commonly used DOA algorithms, such as the SRP-PHAT [2] or the MUSIC [3], on the extracted direct-path bins.

The power steering function for the SRP-PHAT is calculated for all bins in $\Pi$ as

$$P_{\text{SRP-PHAT}}(\Theta) = \sum_{(t,f) \in \Pi} \frac{\left\| \mathbf{x}(t,f)^H \mathbf{g}(f,\Theta) \right\|_2^2}{\left\| \mathbf{x}(t,f) \right\|_2^2}, \quad (7)$$

where $\Theta$ represents the incident angle, and the superscript "H" denotes the conjugate transpose operation of the complex matrix. The MUSIC spectrum is calculated as

$$P_{\text{MUSIC}}(\Theta) = \sum_f \frac{1}{\left\| \mathbf{U_n}(f)^H \mathbf{g}(f,\Theta) \right\|_2^2}, \quad (8)$$

where the matrix $\mathbf{U_n}(f)$ represents the noise subspace assuming a single source, and its columns includes the eigenvectors of the spatial spectrum matrix

$$\mathbf{R}(f) = E\left[ \mathbf{x}(t,f) \mathbf{x}(t,f)^H \right] \quad (9)$$

corresponding to the $M-1$ smallest eigenvalues of a $M$-element array. Note that the expectation in Eq. (9) is estimated by sample average within the set $\Pi$. The angle of the dominant peak is identified as the DOA of the desired speaker.

## 3. Simulations

In this section, the proposed methods are evaluated using a 4-element uniform linear array (ULA) with the inter-microphone distance of 3.5 cm and a 4-element uniform circular array (UCA) with the radius of 3.5 cm. The arrays are positioned in 2 different rooms with parameters shown in Table 1. The room impulse responses (RIRs) are simulated using the image method [19] with room dimensions, array center, and source-array distance, perturbed by 10%. Both directional noise and diffuse noise are considered, and the acoustic noise field generator [20] is utilized to generate the diffuse noise.

Table 1: *Configuration for different rooms.*

| Room | 1 | 2 |
|---|---|---|
| T60 (s) | 0.32 | 0.65 |
| Source-array distance (m) | 3 | 2 |
| Room size (m$^3$) | 7.32×5.5×3 | 5.9×4.2×3.3 |
| Array center (m) | 3, 2.1, 1.2 | 2.5, 1.8, 1.5 |

Table 2: *Configuration for training data generation.*

| Items | Parameter |
|---|---|
| Room size (m$^3$) | [6, 8]×[4, 6]×[2.8, 3.6] |
| Source-array distance (m) | [1.5, 2.5] |
| T60 (s) | [0.16, 2.1] |
| SNR (dB) | [−5, 20] |
| DOA (º) | [−90, 90] |

The TIMIT [21] database is used as the speech source and 18 noises from the Diverse Environments Multichannel Acoustic Noise Database (DEMAND) [22] are used as the noise source, with a sampling rate of 16 kHz, an FFT size of 512 samples, and STFT analysis using a Hanning window with 75% overlap. An analysis frequency range of [1000, 8000] Hz is employed, which leads to a total of 57,600 TF bins for each 2.072 s segment of recordings. $\xi_n$ and $IRM_0$ in Eqs. (2), (3) and (5) are set as 1e$^{-4}$ and 0.5, respectively. The threshold $TH$ in Eq. (6) for each utterance is chosen such that 1000 bins pass the test. The DOA range is −90º to 90º with a 1º resolution.

The multi-task U-net is trained using a single-channel database, with 5300 utterances from TIMIT and 14 different noises from DEMAND randomly chosen to construct the training set. The rest 1000 utterances and 4 noises are randomly divided into the validation set and test set. The configuration of data generation is given in Table 2, and all the parameters are randomly distributed within the labeled upper and lower limits. RIRs from a point source in a room to a microphone are simulated using the image method. We choose the clean utterances from training set, convolve them with the generated RIRs, and then add noise with different SNRs. Overall 70 hours noisy speeches are generated as the training data.

The input log-magnitude spectrograms (256×256 points) to the network are all normalized to [−1, 1]. The mini-batch size of training is 32. The learning rate is set to 1e$^{-4}$ initially, and it will be halved if the loss function of validation set does not decline in 5 consecutive epochs. When the loss function of the validation set does not decline in 30 consecutive epochs, the training will be ended. Our network is trained using the ADAM optimizer with 4 NVidia GTX 1080 Ti GPUs.

### 3.1. Benefit of multi-task learning

A typical test example is presented here to show the benefit of multi-task learning in adverse environment. The test data is generated for the ULA in Room 1 and, with the speaker and the directional noise source placed at 30º and −30º respectively. The SNR is 0 dB.

As shown in the 1.5-1.6 s period of Fig. 3(*b*), a significant amount of noise bins have passed the DPD test if only $IRM_d^*$ is utilized, resulting in an erroneous DOA estimate at the noise direction, as shown in Fig. 3(*e*). In Fig. 3(*d*), it can be seen that most of the noise bins are filtered by the refined DPD test proposed in Sec. 2.3, thus an accurate estimation of the DOA of the desired speaker can be achieved, as shown in Fig. 3(*f*).

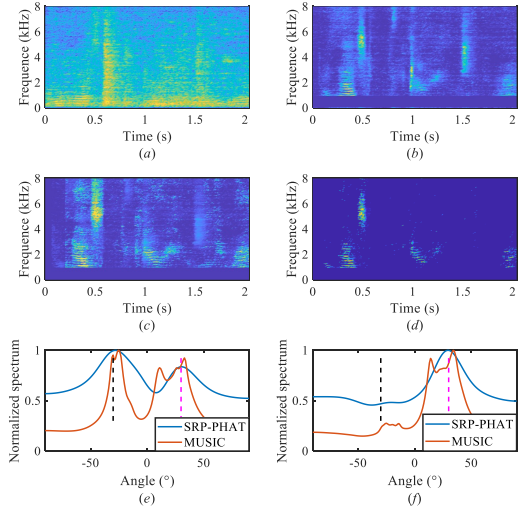

Figure 3: (a) Spectrogram of noisy speech with SNR of 0 dB. (b) TF map of $IRM_d^*$. (c) TF map of $IRM_s^*$. (d) TF map of $IRM_{DPD}$. (e) Normalized spatial spectrum calculated on $IRM_d^*$. (f) Normalized spatial spectrum calculated on $IRM_{DPD}$. The black and magenta dotted lines in (e) and (f) are the DOAs of noise and speech, respectively.

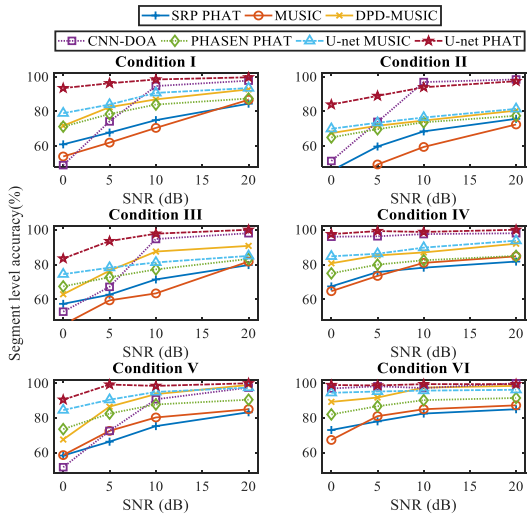

Figure 4: Segment level accuracy of different DOA estimation methods in different conditions.

Table 3: Configuration for different test conditions.

| Condition | Array | Room | Direction-of-arrival | |
|---|---|---|---|---|
| | | | Speaker | Noise |
| I | ULA | 1 | 30º | −30º |
| II | ULA | 1 | 60º | −30º |
| III | ULA | 2 | 30º | −30º |
| IV | ULA | 2 | 30º | Diffuse |
| V | UCA | 2 | 30º | −30º |
| VI | UCA | 2 | 30º | Diffuse |

### 3.2. DOA estimation in reverberant noisy conditions

The performance of the DOA estimation methods is evaluated in terms of segment level accuracy. We consider that the estimate is correct if the difference between the prediction and the true DOA is less than or equal to 5º. Overall, there are 12000 test samples divided into 6 conditions, shown in Table 3. In each condition, there are 2000 noisy utterances with 4 different SNRs, i.e., 0 dB, 5 dB, 10 dB and 20 dB.

Overall 7 different DOA estimation methods are compared, including (i) SRP-PHAT: the conventional SRP-PHAT [2]; (ii) MUSIC: the conventional MUSIC [3]; (iii) DPD-MUSIC: DPD-MUSIC for arbitrary arrays [7]; (iv) CNN-DOA: the CNN based DOA estimation method proposed in [13]; (v) PHASEN PHAT: SRP-PHAT weighted by the amplitude mask estimated using PHASEN [25]; (vi) U-net MUSIC: MUSIC for all bins in Π; and (vii) U-net PHAT: SRP-PHAT for all bins in Π. It should be noted that PHASEN is not proposed for the DOA estimation task but a SOTA deep learning solution of speech enhancement in STFT domain.

Figure 4 presents the segment level accuracy for different methods. From the results, it can be verified that the traditional signal processing based methods, like the SRP-PHAT and the MUSIC, suffer from performance degradation in presence of noise and reverberation [26]. The method of the PHASEN PHAT performs better than traditional methods, but it is still not satisfactory enough. This can be attributed to the fact that the PHASEN is optimized for speech enhancement, not for DOA estimation task. Though the DPD-MUSIC performs better considerably, its performance degrades when SNR becomes lower or the true DOA is closer to the endfire direction of ULA. The performance of the CNN-DOA is close to our proposed method under diffuse noise condition, but seriously degrades at lower SNR with directional noise, which matches well with the results presented in [13]. The U-net PHAT outperforms the U-net MUSIC, because the number of bins passing the DPD test is limited, restricting the performance of MUSIC.

Overall, compared to the other methods, the proposed U-net PHAT method achieves the highest localization accuracy in all the testing scenarios. Its benefit at low input SNRs with directional noise is more remarkable. It also should be noted that the proposed U-net PHAT method can effectively increase estimation accuracy when the expected DOA is close to the endfire direction of the ULA. In addition, the network is trained using a single-channel database, which makes the proposed method easy to be implemented in different arrays. This has been validated by the efficacy of the proposed method on both the ULA and UCA.

## 4. Conclusion

In this paper, we propose a robust DOA estimation method with U-net based extraction of the direct-path TF bins of the target speaker. A multi-task U-net structure is proposed to effectively predict the IRM of both the reverberant speech and the direct speech signal at each TF bin. The training of the network only depends on the input of a single microphone, which makes the proposed method suitable for any array structure. The estimated IRMs are further utilized to refine the direct-path TF bins of the desired target speaker, based on which the DOA is finally estimated by using the common algorithms like the SRP-PHAT. Simulation results validate the benefit of the proposed method especially at low input SNR with directional noise.

## 5. Acknowledgements

This work was supported by the National Natural Science Foundation No. 11874219 of China.